\pdfoutput=1
\documentclass[twocolumn, final]{svjour3}

\usepackage[pdftex,backref]{hyperref}
\hypersetup{
  pdfauthor={Fatemeh Tabatabaei, Olaf Lenz, Christian Holm},
  pdftitle={Simulational study of anomalous tracer diffusion in
    hydrogels}
}

\usepackage{graphicx}
\usepackage{xspace}

\newcommand{\todo}[1]{}

\newcommand{\ie}{\emph{i.e.\/}\xspace} 
\newcommand{\eg}{\emph{e.g.\/}\xspace}
\newcommand{\etal}{\emph{et al\/}\xspace}

\begin{document}

\title {Simulational study of anomalous tracer diffusion in hydrogels}

\author{Fatemeh Tabatabaei \and Olaf Lenz \and Christian Holm}

\institute{
  Fatemeh Tabatabaei \and 
  Olaf Lenz \and 
  Christian Holm 
  \at
  Institut f\"{u}r Computerphysik\\
  Universit\"{a}t Stuttgart\\
  Pfaffenwaldring 27\\
  D-70569 Germany\\
  \email{olenz@icp.uni-stuttgart.de}
}

\date{\today}

\maketitle

\begin{abstract}
  In this article, we analyze different factors that affect the
  diffusion behavior of small tracer particles (as they are used \eg
  in fluorescence correlation spectroscopy (FCS)) in the polymer
  network of a hydrogel and perform simulations of various simplified
  models.  We observe, that under certain circumstances the attraction
  of a tracer particle to the polymer network strands might cause
  subdiffusive behavior on intermediate time scales. In theory, this
  behavior could be employed to examine the network structure and
  swelling behavior of weakly crosslinked hydrogels with the help of
  FCS.
\end{abstract}  

\section*{Introduction}
\label{sec:intro}

Fluorescence Correlation Spectroscopy (FCS) is an experimental
technique that allows to study the diffusion behavior of tracer
particles in a surrounding medium in great detail.

The method is based on detecting the fluctuations of the fluorescent
light intensity in a small and fixed volume element, usually formed by
a laser focus of submicron size.  There is a broad range of application
for FCS, from studying the diffusive dynamics of simple colloids and
polymers to biomolecules as well as hydrogels.
\cite{wink07a,lumma03a,nandi08a,bacia04a,cherdhirankorn09a,cherdhirankorn09b}

To describe the diffusion behavior of tracer particles, one typically
uses the mean square displacement (MSD) $\langle |\Delta\vec{r}(t)|^2
\rangle$, where $\vec{r}(t)$ is the average displacement of the
particle after a time $t$.  Often, the MSD of a particle behaves as
\begin{equation}
\langle |\Delta\vec{r}(t)|^2 \rangle \propto t^\alpha
\end{equation}
where $\alpha$ is the diffusion exponent. The case where the MSD
scales linearly with time ($\alpha = 1$) is called \emph{normal
  diffusion}, and is the most common case in physical processes, for
example for particles that perform a simple random walk. A value of
$\alpha \ne 1$ signifies \emph{anomalous diffusion}, with the special
cases of \emph{sub-diffusion} ($\alpha < 1$) and \emph{superdiffusion}
($\alpha > 1$).  During the last decades, there has been considerable
interest in anomalous diffusion, and such behavior has been found in
many systems. \cite{gallo03a,diet08a,wong04a,tolic04a,banks05a,szymanski09a,saxton96a}

Fytas \etal have employed the FCS technique to study the diffusion of
tracer particles in a hydrogel \cite{cherdhirankorn09a}. They found
evidence, that under certain circumstances, tracer particles in a
hydrogel network exhibit anomalous diffusion behavior.\cite{fytas08a}

The goal of this work is to approach the problem from a theoretical
point of view, to explain how anomalous diffusion behavior can occur
in such a system, and the different processes that play a role in the
diffusion.  From the theoretical considerations, it is shown that FCS
may be a valuable tool for examining the network structure and
swelling behavior of real hydrogels.

Sprakel \etal\cite{sprakel07a} have studied the diffusion of colloidal
particles in polymer networks via photon correlation spectroscopy.
They observe, that colloids that can bind to the polymer strands of
the network show subdiffusive behavior. They explain this by the
observation, that colloids that are bound to multiple polymer strands
behave as though they are a part of the strand themselves, and thus
show the Rouse-dynamics of a polymer segment.  The authors develop a
simple model that explains the observations.

In our studies, the FCS tracer particles are smaller than the
colloidal particles in the above work, and they are assumed to be
significantly smaller than a network mesh cell.  Furthermore, they do
not bind to the polymer strands, although there might be attraction
between the tracer and the network strands.  To our best knowledge,
there are no theoretical studies of this situation.  In this work, we
have focused on the behavior of tracer particles that are attracted to
the polymer network and can slide along single strands.  This gives
rise to a number of interesting phenomena.  The following thoughts are
closely related to the biophysical problem of proteins sliding along
DNA strands. \cite{hu06a,berg81a}

\section*{Processes affecting the diffusion behavior}

\begin{figure}[htbp] 
  \centering
  \centering
	\includegraphics[width=0.8\linewidth]{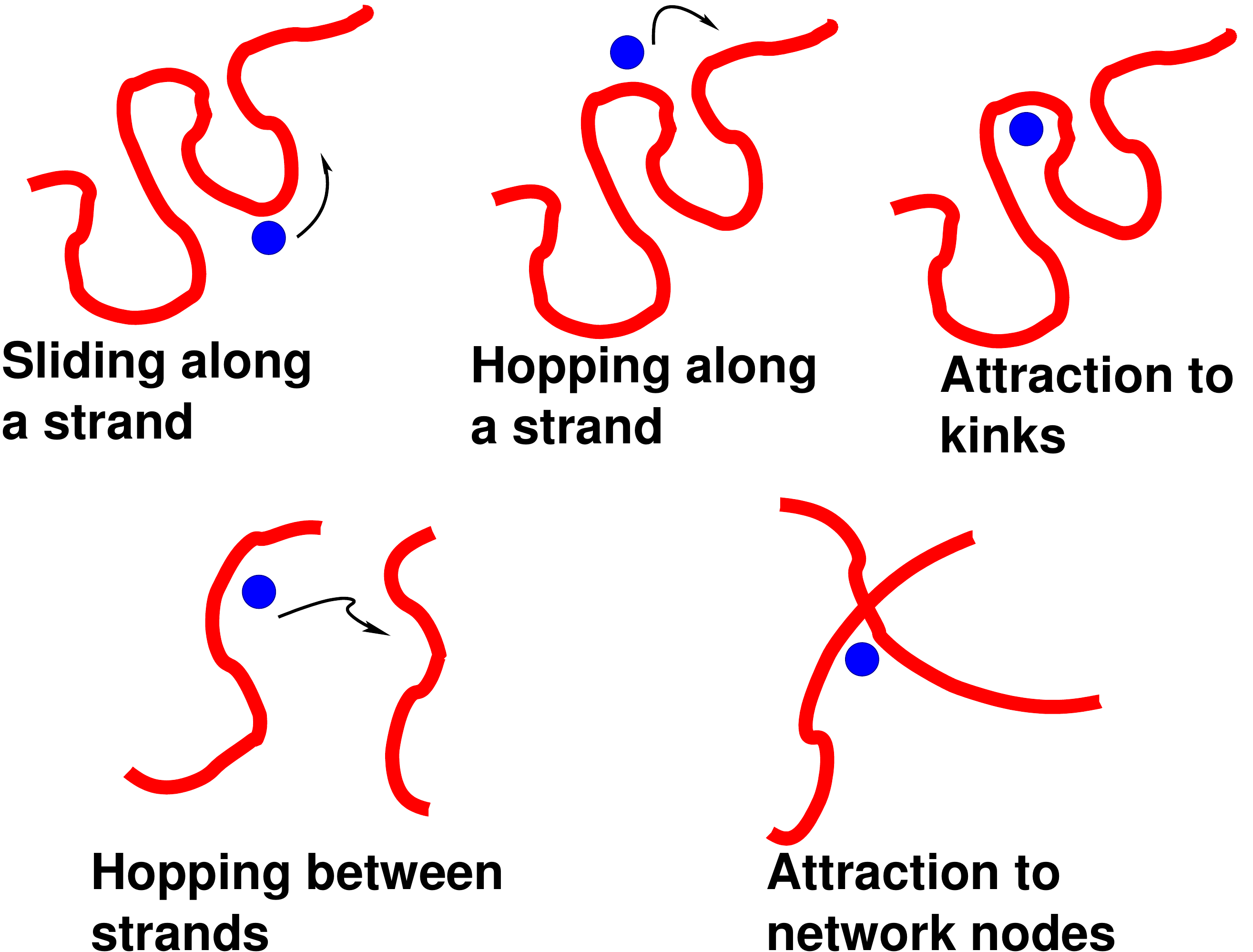}
  \caption{Schematics of the different processes possibly involved in
    the diffusion of a small, non-specifically bound tracer particle
    in a hydrogel.}
  \label{fig:processes}
\end{figure}

Several processes that can theoretically influence the diffusive
behavior of the tracer particles in the hydrogel's polymer network can
be identified (see Figure \ref{fig:processes}).

\subsection*{Sliding}
From the point of view of this work, the most interesting process is
the process of tracer particles \emph{sliding along a polymer strand}.
While tracer particles that are non-specifically bound to a polymer
strand can not easily be detached from the strand by the random kicks
of the solvent particles, they can still perform a one-dimensional
random walk along the strand.  In that case, the particles can be
considered as ``diffusing'' along the contour of the polymer.  In this
random walk, the mean contour length $c$ that the tracer has covered
after the time $t$ behaves as
\begin{equation}
  \label{eq:c_t}
  \langle c \rangle \propto t^\frac{1}{2}.
\end{equation}

On the other hand, as it is well known since Flory's work
\cite{flory53a}, the contour of a relaxed polymer strand itself can be
described as a random walk (RW) or a self-avoiding random walk (SAW)
in three-dimensional space.  It follows, that the MSD of a monomer
$\langle |\Delta\vec{r}(t)|^2 \rangle$ to another monomer with a
distance $c$ along the contour behaves as
\begin{equation}
  \label{eq:msd_c_rw}
  \langle |\Delta\vec{r}(t)|^2 \rangle \propto c
\end{equation}
in the case of the random walk, or as 
\begin{equation}
  \label{eq:msd_c_saw}
  \langle |\Delta\vec{r}(t)|^2 \rangle \propto c^\frac{6}{5}.
\end{equation}
in the case of the self-avoiding random walk.  Substituting equation
\ref{eq:c_t} into equations \ref{eq:msd_c_rw} or \ref{eq:msd_c_saw},
this results in the MSD $\langle |\Delta\vec{r}(t)|^2 \rangle$ of the
tracer particle behaving like
\begin{equation}
  \label{eq:msd_rw}
  \mathrm{RW:}\> \langle |\Delta\vec{r}(t)|^2 \rangle \propto t^\frac{1}{2} 
\end{equation}
or
\begin{equation}
  \label{eq:msd_saw}
  \mathrm{SAW:}\> \langle |\Delta\vec{r}(t)|^2 \rangle \propto t^\frac{3}{5}.
\end{equation}
\ie the diffusion exponents are $\alpha = \frac{1}{2}$ (RW) and
$\alpha = \frac{3}{5}$ (SAW) respectively, which denotes strongly
subdiffusive behaviour.

Equations \ref{eq:msd_rw} and \ref{eq:msd_saw} are only valid in the
case of completely relaxed polymer coils.  In the other extreme case,
when the polymer is completely stretched and rod-like, the spatial
distance of two monomers is identical to the contour distance of the
monomers.  Tracer particles sliding along such a rod exhibit normal
diffusive behavior ($\alpha = 1$).  Consequently, when stretching a
polymer and thus crossing over from relaxed coil states to stretched
rod-like polymer states, a transition from subdiffusive to normal
diffusive behavior must take place.  Note, that in both cases the
diffusion exponent applies to all length and time scales.

When the polymer strand is neither fully stretched nor fully relaxed,
the polymer structure is becoming more complex.  Stretching the
polymer chain does not affect the polymer structure on all length
scales equally.  The structure of the stretched polymer on short
length scales will be unaffected by the stretching and retain the
random-walk or self-avoiding random walk behavior, while on longer
length scales, the polymer is stretched and becomes rod-like.  A
simple description of this behavior is given by Pincus' blob picture.
In that picture, a polymer chain under tension can be seen as a chain
of \emph{blobs}, each of which contains a distinct segment of the
polymer.  While the chain of blobs behaves rod-like, each of the blobs
itself consists of a polymer segment that can still be described well
by a RW or SAW.  The size of the blobs is determined by the
stretching; the larger the amount of stretching, the smaller the
blobs. \cite{pincus76a,degennes79a}

When applying this picture to the behavior of the tracer particles
that slide along the polymer contour, the MSD of the tracer particles
can be expected to show a more complex behavior at different time
scales, which would be characterized by varying diffusion exponents at
different time scales.  At small time scales, the tracer particles
probe the random-walk-like internal structure of the Pincus blobs, and
the subdiffusive behaviour described above should be observable, with
small exponents down to $\alpha = \frac{1}{2}$ (RW) and $\alpha =
\frac{3}{5}$ (SAW), respectively.  At longer time scales, the tracer
particles probe the stretched structure of the sequence of Pincus
blobs, which leads to normal diffusive behavior and an exponent of
$\alpha = 1$.

However, formally, the diffusion exponent $\alpha$ is only defined
asymptotically for large timescales, therefore it is not very useful
in our case.  To be able to distinguish more details of the diffusion
behavior at different time scales, we instead examine the
time-dependent exponent $\alpha(t)$
\begin{equation}
  \alpha(t) = \frac
  {\mathrm{d}\log \langle |\Delta\vec{r}(t)|^2 \rangle}
  {\mathrm{d}\log t}
  \label{eq:alphat}
\end{equation}

In the case of hydrogels, the degree of stretching of the polymer
strands is directly influenced by the swelling ratio of the hydrogel.
While in a dry or only slightly swollen hydrogel, the strands can be
expected to be mostly relaxed, the strands must be stretched for
strongly swollen hydrogels.  This means, that the diffusion behavior
of non-specifically bound tracer particles should exhibit the
characteristics described above and should change accordingly when the
gel is swollen.

If this effect occurs in the parameter range of real hydrogels and is
not dominated by other processes that influence the diffusion behavior
of the tracer particles, it could provide a nice way to explore the
structure of the polymer network and the swelling of hydrogels via
FCS.  One should note, however, that this method only works if the
average length of the polymer segments (which corresponds to the
average network mesh cell size) and the average time required to slide
along these segments are larger than the minimal time and length
resolution of the FCS technique.  For most real hydrogels, this is not
the case.  Only in weakly crosslinked hydrogels, it can be expected
that the mesh size comes into regions where this effect might be observable.

\subsection*{Hopping}

The above discussion only holds for an ideal model where the tracer
particles are sliding along the contour of the polymer.  In reality,
however, a number of other processes play a role in the diffusion
behavior of tracer particles.

When the tracer particles are not very strongly attracted or bound to
the polymer strand, they can be ripped off by thermal motion and
diffuse freely in the solvent until they are adsorbed again by the
same or another polymer strand.  This process of escape and
readsorption is referred to as \emph{hopping} throughout this work.
There are actually two hopping processes; the tracer particle can be
reabsorbed by the same network strand as it was ripped off, which is
referred to as \emph{hopping along a strand}, or it is absorbed by
another polymer strand, which is called \emph{hopping between
  strands}.

When hopping occurs, the tracer particle diffuses freely in the
solvent.  Compared to the sliding process depicted above, it can be
expected that free diffusion has a significantly larger diffusion
constant than the sliding process and a diffusion exponent of $\alpha
= 1$, as it is a normal diffusive process.  Since the diffusion of the
hopping particles is much faster than of the sliding particles, there
is the danger that it dominates the overall diffusive behavior and
effectively hides the anomalous behavior of the sliding process.  The
relative importance of hopping towards sliding is controlled by the
following parameters:
\begin{itemize}
\item The strength of the attractive interaction governs the hopping
  rate: the stronger the attraction, the lower the probability for the
  tracer to escape from the polymer strand, and the higher the
  probability to get readsorped.  This parameter affects both
  hopping processes.
\item The density of the polymer strands directly controls the mean
  free path that the tracer can cover before it gets absorbed by
  another polymer strand, \ie it influences the effect of hopping
  between strands.
\item The spatial correlations between different parts of the polymer
  strands are unequal in the coiled and stretched states.  In the
  coiled state, a tracer particle is more likely to be readsorped onto
  another part of the same polymer strand, as in the stretched state.
  This means that the process of hopping along a strand is suppressed
  by stretching the strand, while hopping between strands should be
  mostly unaffected.
\end{itemize}

\subsection*{Sticking}

It is likely that network nodes or kinks in a polymer strand, where
tracer particles can synchronously interact with multiple polymer
segments further affect the diffusive behavior.  The strong attraction
of the tracer particles to the polymer strand at these locations might
immobilize the tracer particles, which reduces the overall diffusion
rate.\cite{saxton96a}

\section*{Models}

Starting from the assumption that tracer particles can slide along the
polymer contour and therefore show subdiffusive behavior, the goal of
this work was to study different influence factors on this behavior
with the help of simulations, to quantify them and their relation to
each other, and to distinguish their different origins.

To this end, we have developed a series of coarse-grained, simplified
diffusion models of the system where we successively add details and
allow more and more of the processes depicted above to happen.

All of the diffusion models used fixed polymer conformations, \ie the
polymer did not move throughout the diffusion simulation.  This
simplifies the models to a great extent, as it takes out the
influencing factors of the various parameters of the polymer
simulation, at the cost of not describing correctly the influence of
the polymer dynamics on the tracer particles.  We believe that the
dynamics of the polymer will not influence the diffusive behavior of
the tracer particles to a great extent, although it might shift some
of the results obtained in this work.

Furthermore, in the models used in this work, we did not consider the
case of a polymer network with connections between different polymer
strands, but we always used a single polymer strand of infinite
length.  Consequently, we did not observe the effect of network nodes
on the diffusion behavior.  We plan to consider both the effect of the
dynamics of the polymer as well as the effect of a polymer network
with network nodes in our future work.

\subsection*{Polymer model}

All of the diffusion models used conformations of an coarse-grained
bead-spring polymer with different degrees of stretching.  The
conformations have been generated via a standard Langevin dynamics
simulation of a three-dimensional model polymer system with periodic
boundary conditions.

\begin{figure}[htbp] \centering
  \centering
	\includegraphics[width=0.35\textwidth]{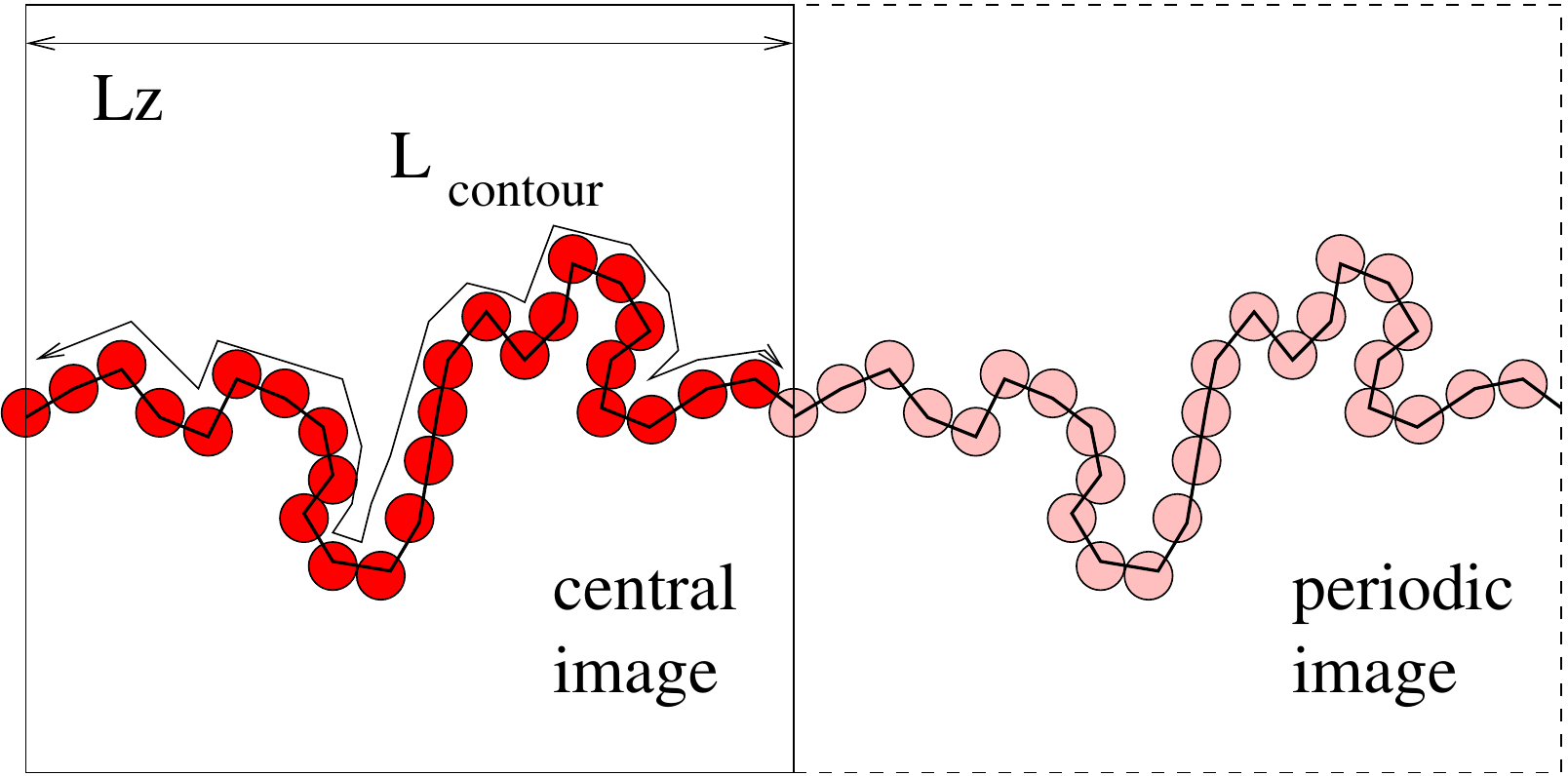}
  \caption{Sketch of the system with a stretched periodically repeated
    polymer.  The degree of stretching is defined by the number of
    monomers $N$ and the system size along the axis of stretching
    $L_z$.}
  \label{fig:stretched_polymer}
\end{figure}

The polymer consisted of $N=200$ beads (or ``monomers''), each with
mass $m=1$, that were bound to each other using a harmonic spring
potential to form a chain.
\begin{equation}
  \label{eq:harmonic}
  V_\mathrm{harmonic}(r) = \frac{1}{2} K \left( r - r_0 \right)^2
\end{equation}
The parameters were chosen to be $r_0=0.5$ and $K=10.0$.

To generate a polymer of seemingly infinite length with various
degrees of stretching, the polymer was periodically repeated along the
$z$-axis of the system, \ie the last monomer of the chain was bound to
the first monomer of the chain over the periodic boundary (see figure
\ref{fig:stretched_polymer}). Note that this does \emph{not} mean that
any of the beads was fixed in any spatial dimension throughout the
simulations.

In this model, the degree of stretching depends on the number of
monomers $N$ and the size of the system along the $z$-axis $L_z$.
Given a value of $L_z$, the system lengths $L_x = L_y$ could be used
to control the monomer density $\rho$ of the system.

The unusual setup of the periodically repeated polymer also had some
effects on the Pincus blob picture mentioned above.  On the one hand,
the system size $L_z$ limited the maximal size of a Pincus blob.  As
the various periodic images were always aligned chain-like, the
polymer always displayed a rod-like structure for length scales larger
than $L_z$.  Accordingly, the time-dependent diffusion exponent
$\alpha(t)$ always approached $1$ when the tracer particles diffused
further than this length scale.  It should be noted that this was an
artifact of the simulation setup, which limited the maximal Pincus
blob size.  In real systems with fully relaxed chains, it should in
theory be possible to see the subdiffusive behavior on larger length
and time scales.  On the other hand, when $L_z$ was chosen smaller
than the expected size of a completely relaxed polymer coil (given by
Flory's argument), this corresponds to an unphysical compression of
the polymer in this direction.  Therefore, such states were avoided
throughout this work.

Two variants of the polymer model were simulated. In the first
variant, no excluded volume interaction between monomers was used,
therefore the conformations in the coiled states correspond to a
random walk (RW).  The RW conformations were used only in conjunction
with the strongly bound tracer model described below to verify the
theoretical considerations above.  In the second variant, a repulsive
Weeks-Chandler-Anderson (WCA) interaction between the monomers was
used to model the excluded volume interaction
\cite{allen87a,padding06a}, which is identical to the repulsive core
of the Lennard-Jones interaction cut off at its minimum (equation
\ref{eq:WCA}, $\varepsilon=1.0$, $\sigma=1.0$).  In this case, the
conformations corresponded to a self-avoiding random walk (SAW)
\cite{milchev02a}.
\begin{equation}
  \label{eq:WCA}
  V_\mathrm{WCA}(r) = \Biggl\{
    \begin{array}{ll}
      4\varepsilon((\frac{\sigma}{r})^{12}
      - (\frac{\sigma}{r})^6) 
      & \mathrm{, if~} r < (2\varepsilon)^\frac{1}{6}\\
    0 & \mathrm{, otherwise}\\
    \end{array}.
\end{equation}

The system was simulated with help of the simulation package
\textsc{ESPResSo} \cite{limbach06a}, using the Velocity Verlet
algorithm with a time step of $\tau=0.005$ and a Langevin thermostat
with temperature $T=1.0$ and a friction parameter $\gamma=0.5$.
Various simulation runs with up to $50 \times 10^6 \tau$ in total and
$500,000 \tau$ between single polymer conformations were performed, to
obtain $100$ equilibrated, statistically independent polymer
conformations for each different value of the number of monomers $N$,
degree of stretching (\ie $L_z$) and polymer density $\rho$.  Note
that $\rho$ was always chosen small enough so that the different
periodic images of the polymer strand in $x$ or $y$ direction did not
interact with each other.  For the given parameters, the expected
end-to-end distance (and hence the value of $L_z$) of the RW polymer
is about $12$, while for the SAW polymer it is about $20$ (given the
measured average bond length of about $0.7$).  This means that the
polymer conformations at $L_z = 15$ (RW) resp. $L_z = 30$ (SAW)
correspond to mostly relaxed polymer chains, while polymer
conformations at $L_z = 100$ (RW and SAW) are strongly stretched.

\subsection*{Strongly bound tracer model}
In the simplest diffusion model, only the process of the tracer
particle sliding along the polymer contour was considered.  To this
end, the frozen polymer conformations were used as a basis for a
one-dimensional random walk along the polymer contour, where the
tracer particles were thought to be strictly bound to the polymer and
could not escape.

To do that, a random monomer from each conformation was used as a
first binding location of the virtual, strongly bound tracer particle.
When advancing a ``timestep'', either the same, the next or the
previous monomer along the chain was randomly chosen as the next
binding location of the virtual particle, to model the diffusion of
the tracer along the polymer.  After $t$ timesteps, the MSD of the
binding location was measured to determine the spatial diffusion
behavior of the virtual tracer.

The model was simulated for the various conformations obtained in the
polymer model, and the dependence of the diffusive behavior of the
virtual tracer on the degree of stretching (\ie $L_z$) and the chain
length $N$ was studied.  As the virtual tracer particle could not
leave the single infinite strand, and the different periodic images of
the polymer strand did not interact with each other, the monomer
density $\rho$ does not play a role in this model.

The results were used to verify the theoretical considerations from
above.  Therefore, the model was simulated both with the RW polymer
conformations as well as the SAW polymer conformations.

\subsection*{Freely diffusing tracer particles with nearest-neighbor
  interaction}

On the next level of model refinement, the influence of the hopping
process onto the diffusive behavior of the tracer particle was
studied.  The model was refined, and a simple model of a single tracer
particle was introduced.  The tracer particle was represented by a
bead of mass $m=1$ that interacted with the (fixed) polymer beads via
a Lennard-Jones interaction (equation \ref{eq:LJ}, $\sigma=1.0$,
$r_\mathrm{cut}=2.0$, $\varepsilon$ varied).  As the polymer beads
were fixed in space, this can also be interpreted as the diffusion of
the tracer particle in a system of fixed obstacles.
\begin{equation}
  \label{eq:LJ}
  V_\mathrm{LJ}(r) = \Biggl\{
    \begin{array}{ll}
      4\varepsilon((\frac{\sigma}{r})^{12}
      - (\frac{\sigma}{r})^6)
      & \mathrm{, if~} r < r_\mathrm{cut}\\
    0 & \mathrm{, otherwise}\\
    \end{array}.
\end{equation}

On this refinement level, we still wanted to avoid the effect of the
tracer particle getting stuck at kinks of the polymer strand, where
the bead would feel a significantly deeper potential well than close
to a stretched piece of the polymer.  To achieve that, we have
modified the interaction such, that the tracer bead only interacted
with the single monomer on the polymer strand that was closest to the
tracer particle.  We call this the \emph{nearest-neighbor
  interaction}.  This simple trick prevented the tracer particle from
getting stuck at kinks, while it still felt the attraction of the
strand.

The most important parameter introduced by this model was the
parameter $\varepsilon$ of the interaction, which determined the
strength of the attraction between the tracer particle and the polymer
and consequently the rate of the hopping process.  Compared to a
system with an all-neighbor LJ interaction, the value of $\varepsilon$
had to be chosen much larger, as in this model the tracer particle
always only feels the attraction of a single monomer as opposed to
several monomers at once when being close to the strand in a normal
Lennard-Jones all-neighbor interaction.  Another parameter that
influenced the behavior of this model was the monomer density $\rho$,
which mostly controlled the rate of the hopping between polymer
strands.

Langevin MD simulations (time step $\tau=0.01$, temperature $T=1.0$,
friction $\gamma=0.5$) of the tracer were performed.  The MSD and the
time-dependent exponent $\alpha(t)$ of the tracer diffusion was
measured.  The model was studied for various SAW polymer conformations
at different degrees of stretching $L_z$ and different chain lengths
$N$, as well as at different monomer densities $\rho$.  The RW polymer
conformations were not considered in this model.

It is important to note that there is no relation between the time
scale in this model and the time scale in the strongly bound tracer.
While the time scale in the strongly bound tracer model was a pseudo
time scale in units of the time that the tracer particle needs to
diffuse to the next monomer in the polymer chain, the time unit in
this model is determined by the LJ parameters $\varepsilon$ and
$\sigma$.

\subsection*{Freely diffusing tracer particles}
The last model considered in this work was mostly identical to the
previous model with nearest-neighbor interaction, but used a
conventional all-neighbor Lennard-Jones interaction (Equation
\ref{eq:LJ}), so that the tracer particle interacted with all polymer
beads within cutoff range.  Thus, the effect of the tracer particle
getting stuck to kinks was included in this model.

The simulation parameters are similar to the previous model.  The only
difference is the depth of the Lennard-Jones potential $\varepsilon$,
which had to be chosen significantly lower in this model to exhibit a
comparable effect.

\section*{Results}
\subsection*{Strongly bound tracer model}
\begin{figure}[htbp]  
  \centering
  \centering
	\includegraphics[width=0.95\linewidth]{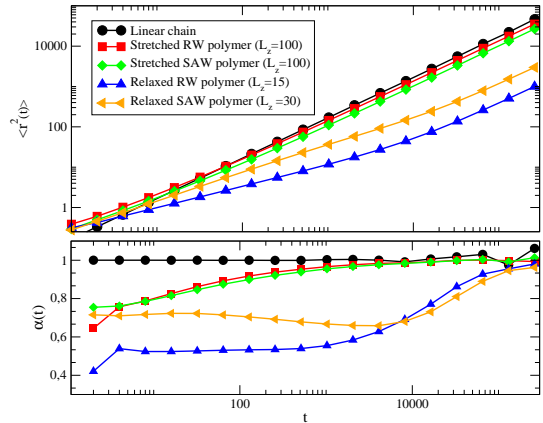}
  \caption{Top: MSD of the strongly bound tracer model for both polymer
    model variants (RW and SAW) at two different degrees of
    stretching. Bottom: Time-dependent exponent of the curves in the top
    plot.}
  \label{fig:m1_msd}
\end{figure}

The upper subplot of figure \ref{fig:m1_msd} shows the MSD of the
tracer particles in the strongly bound tracer model, for both RW and
the SAW polymers as well as for two different representative values of
stretching of the chain, while the lower subplot shows the
time-dependent exponent $\alpha(t)$ in the same cases.  While for the
relaxed polymer conformations $L_z$ was close to the expected
end-to-end distance ($L_z = 15$ in the RW and $L_z = 30$ in the SAW
case), for the stretched conformations it was rather large with $L_z =
100$.

As expected, the plots show a constant diffusion exponent
$\alpha(t)=1$ for a linear chain of monomers, while for the polymer
chains it varies between $\frac{1}{2} < \alpha(t) < 1$ for the polymer
in the RW case and $\frac{3}{5} < \alpha(t) < 1$ in the SAW
case. Interestingly, all of the curves for RW and SAW polymers alike
show values of the time-dependent exponent $\alpha(t)$ at small time
scales that are significantly less than $1$, and approach a value of
$1$ at longer time scales.  This behavior is consistent with the
picture of the strongly bound tracer particle in a polymer that
consists out of Pincus blobs -- at short time scales, the tracer
probes the coiled structure of single Pincus blobs and exhibits
subdiffusive behavior, while at longer time scales, the tracer
particles feel the linear alignment of multiple Pincus blobs, which
results in normal diffusive behavior.

In the case of the stretched polymers (both RW and SAW), the
time-dependent exponent quickly approaches a value of $1$ within a few
orders of magnitude of the time scale, while in the case of the
relaxed polymers, it stays at values of $\alpha(t) \approx 0.5$ (RW)
or $\alpha(t) \approx \frac{3}{5}$ (SAW) over many orders of
magnitude.  This is also well described by the Pincus blob picture,
which predicts that even while the overall structure of the polymer is
unaffected by stretching the polymer, the single blobs are largest for
relaxed polymer coils, while for higher degrees of stretching the
blobs become smaller.  Note, that even in the maximally relaxed
polymer coils in our model the time-dependent exponent $\alpha(t)$ has
to approach a value of $1$ for long time scales $t$, as the Pincus
blob is limited by the system size $L_z$ as described above.
 
\subsection*{Freely diffusing tracer particles with nearest-neighbor
  interaction}

\begin{figure}[H!tbp] 
\centering
	\includegraphics[width=0.95\linewidth]{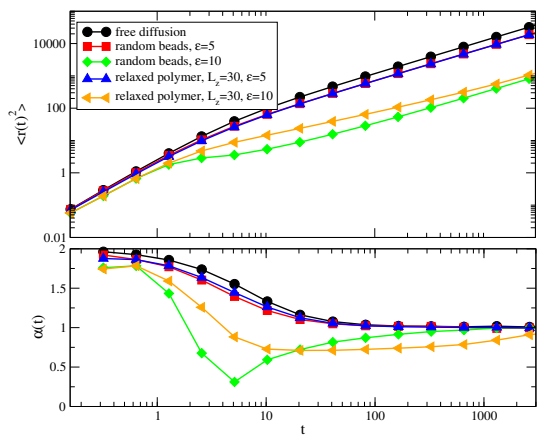}
\caption{MSD and time-dependent exponent of the freely diffusing
  tracer model with nearest-neighbor interaction in a system without
  any obstacles (``free diffusion''), in a system of random beads at
  two different interaction strengths, and in a system with a relaxed
  polymer chain at two different interaction strength.  The density of
  the random beads was chosen to match the polymer results as
  described in the text.}
  \label{fig:m2_var_systems}
\end{figure}

In order to investigate whether it is at all possible to observe an
effect of the polymer structure on the diffusive behavior of the
tracer particles in the tracer model with nearest-neighbor
interaction, we have performed a few simulations of systems that
contained randomly distributed beads as obstacles.  If the polymer
structure has an effect, the diffusion behavior should show distinct
differences in such a system when compared to a system of random
beads.  To be able to do such a comparison, it is important to note
that the beads in the chain have a significant overlap, while random
beads do not.  Therefore, the volume fraction of the system where the
tracer particle interacts with an obstacle bead is significantly
larger in the case of the random beads than in the case of the polymer
chain at similar obstacle bead density.  To be able to see whether the
actual structure of the polymer has a distinct influence on the
diffusive behavior, it is therefore necessary to match the volume
affected by the obstacle beads rather than the density.

The MSD and time-dependent exponent of tracer particles of different
systems of random beads and a relaxed polymer chain are plotted in
Figure \ref{fig:m2_var_systems}.  These plots differ significantly
from the corresponding plots in the strongly bound tracer model.  The
most obvious difference is that at very short time scales of $t < 1$,
the MSD seems to exhibit strong superdiffusion with an exponent of up
to $2$, that was not visible in the strongly bound tracer model.  This
behavior is an artifact of the tracer MD simulation; even in the case
of free diffusion, on these short time scales, the tracer particles
propagated freely and exhibited ballistic motion with a diffusion
exponent of $2$, before the Langevin thermostat destroyed the velocity
autocorrelation of the tracer particles and the normal diffusion sets
in.  This effect is of no interest to this work, therefore none of the
figures following figure \ref{fig:m2_var_systems} show this regime.

At relatively low interaction strength $\varepsilon = 5$, only a very
minor effect of the random obstacle beads on the diffusive behavior is
observed, and the MSD plot shows the same two clearly distinct regimes
-- a short-time ballistic regime, and a longer-time normal diffusive
regime.  The differences can be attributed to the beads acting as
simple obstacles that hinder the diffusion of the tracer and thus
reduce the diffusion constant.  This also holds in the case of the
relaxed polymer chain at low interaction strength, where the MSD is
virtually identical to the random bead system.  No effect of the
polymer structure can be recognized.

When the interaction strength is increased, the effect of the random
beads hindering the diffusion becomes more pronounced, up to a level
where the normal diffusion is so slow that between the ballistic and
normal-diffusive regime a short intermediate region appears, where the
time-dependent exponent drops below $1$.  This region should not be
interpreted as a real subdiffusive regime, instead it is more of an
artifact caused by the combination of two normal diffusion processes
with very different length and energy scales: the diffusion within the
potential of a single bead is very fast, but the tracer is strongly
bound to the bead (``rattling in a cage'').  When the tracer particle
escapes, it can diffuse fast and freely until it is caught by another
attractive bead (``hopping between cages'').  Looking at the overall
diffusion, this results in the observed behavior with the
pseudo-subdiffusive intermediate region.

In the case of the relaxed polymer chain at high enough interaction
strength, the effect is different.  As in the case of the random
beads, an intermediate region occurs between the ballistic and
normal-diffusive regimes, where the time-dependent exponent is below
$1$.  In contrast to the random bead system, however, $\alpha(t)$ is
mostly constant and below $1$ for several orders of magnitude, so that
we can speak of a real subdiffusive regime. 

When examining the MSD plot in more detail, it can be observed that in
the case of the relaxed polymer chain, this subdiffusive regime spans
time scales of up to at least $1000$ time units and the corresponding
length scale of up to a few tens of bead diameters, indicating that
the tracer bead really slided along the polymer contour for a few tens
of monomers.

\begin{figure}[htb] 
\centering
\includegraphics[width=0.95\linewidth]{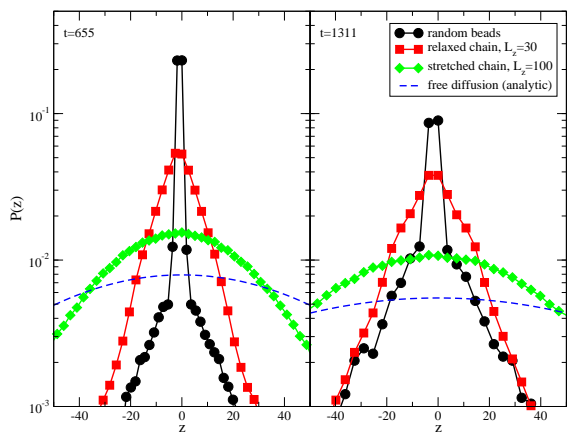}
\caption{Probability distribution of the tracer displacement in $z$
  direction at two different times.  Other parameters: interaction
  strength $\varepsilon=10$, density $\rho=0.0084$.  The free
  diffusion case was computed analytically using Equation
  \ref{eq:px}.}
  \label{fig:m2_dist}
\end{figure}

Plotting the MSD and time-dependent exponent is very useful to look at
the overall diffusion behavior.  A more detailed way to visualize what
is going on in a system is to plot the displacement probability
distribution $P(r)$ at different times.  As we are mainly interested
in the diffusion of the tracer particles along the $z$-axis of the
system, Figure \ref{fig:m2_dist} depicts the displacement probability
distribution $P(z)$ for different systems at different time scales in
the intermediate regime where the tracer particles exhibit
subdiffusion in some of the systems.  In the case of free diffusion,
we can easily compute the displacement distribution function
analytically as in Equation \ref{eq:px}.\cite{benavraham00a,klages10a}
When plotting the measured displacement probability in the free
diffusion case, it is indistinguishable from the analytically derived
function.
\begin{equation}
  P(z,t)=\frac{1}{\sqrt{4\pi Dt}}\exp\left(\frac{-z^2}{4Dt}\right).
  \label{eq:px}
\end{equation}

In the plot, we can discern the qualitative differences between the
various systems.  While the free diffusion case shows Gaussian
behavior on all time scales, in the other systems we can observe a
more complex behavior.  The distributions in the random bead system
are characterized by a sharp peak at $z=0$ with a very small width and
lower, much broader shoulders. This behavior is consistent with a
superposition of a sharp peak and a broader Gaussian, where the peak
originates from the tracer particles being caught in the potential of
a random bead, while the smaller but significantly broader Gaussian
comes from the free diffusion of particles that have escaped the bead
potentials.  The distributions measured in the polymer systems look
distinctively different from both of these extremes.  None of the
curves looks like a clear superposition of two Gaussians, but none of
them really looks like a Gaussian either.

\begin{figure}[hbtp] 
  \centering
	\includegraphics[width=0.95\linewidth]{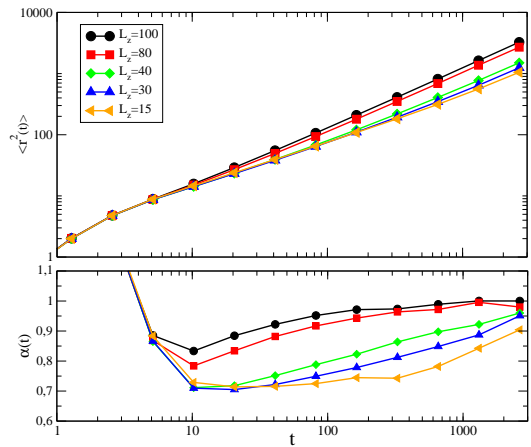}
  \caption{MSD and time-dependent exponent of the freely diffusing
    model tracer with nearest-neighbor interaction for different
    degrees of stretching of the polymer at constant density
    $\rho=0.0084$ and interaction strength $\varepsilon=10$.}
  \label{fig:m2_msd}
\end{figure}

To determine the effect of the degree of stretching of the polymer on
the diffusion behavior, we have performed a number of simulations for
varying degrees of stretching at constant density $\rho=0.0084$ and
constant interaction strength $\varepsilon=10$ (Figure
\ref{fig:m2_msd}).  We observe, that the size of the intermediate
regime seems to depend on the degree of stretching of the attractive
polymer chain in the system; for mostly stretched chains, the
normal-diffusive regime is restored after a few orders of magnitude of
the time scale, while for relaxed chains, the MSD shows subdiffusive
behavior for many decades.  In the most relaxed chains, the
normal-diffusive regime is restored at about the length scale of
$L_z$, which corresponds to the fact that the Pincus blob size is
limited by $L_z$.  Therefore, this behavior is again well consistent
with the Pincus blob picture.

\begin{figure}[htbp] 
  \centering
  \includegraphics[width=0.95\linewidth]{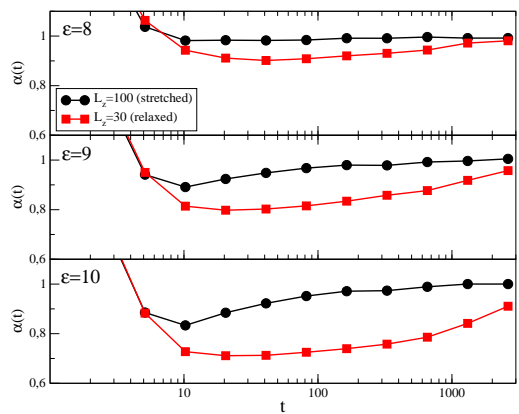}
  \caption{Time-dependent exponent for two different degrees of
    stretching ($L_z=30$ and $L_z=100$) at different values of the
    interaction parameter $\varepsilon$ and constant monomer density
    $\rho=0.0084$.}
  \label{fig:m2_var_epsilon}
\end{figure}
\begin{figure}[htbp] 
  \centering
  \includegraphics[width=0.95\linewidth]{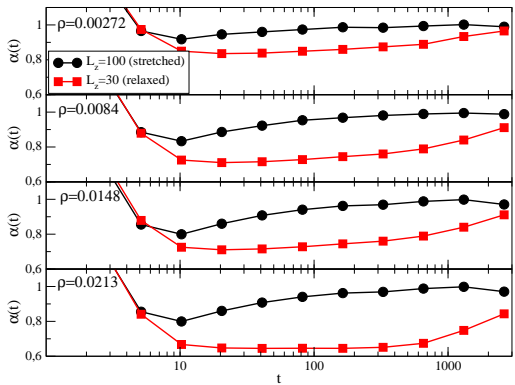}
  \caption{Time-dependent exponent for two different degrees of
    stretching ($L_z=30$ and $L_z=100$) at different values of the
    monomer density $\rho$ and constant interaction parameter
    $\varepsilon = 10$.}
  \label{fig:m2_var_rho}
\end{figure}

Further simulations have been performed for varying values of the
interactions strength $\varepsilon$ (Figure \ref{fig:m2_var_epsilon})
and the monomer density $\rho$ (Figure \ref{fig:m2_var_rho}), to be
able to quantify the influence of these parameters on the diffusion
behavior.  From the plots, it can be concluded that lowering the
interaction strength $\varepsilon$ suppresses the effect of the
polymer chain, as does lowering the density $\rho$.  As discussed in
the introduction, this can be traced to the influence of the hopping
process which dominates the diffusion along the polymer.  Although the
qualitative effect of varying the interaction strength $\varepsilon$
(Figure \ref{fig:m2_var_epsilon}) is similar to varying the density
$\rho$ (Figure \ref{fig:m2_var_rho}), quantitatively the interaction
strength has a significantly higher impact on the subdiffusive regime,
where varying $\varepsilon$ by $20\%$ has a similar effect to varying
$\rho$ by factor $5$.  We interpret this along the lines of what was
discussed in the introduction -- while varying $\rho$ only has an
effect on the less important process of hopping between polymer
strands, the interaction strength has an effect on all hopping
processes, both on the more important process of hopping along the
same strand as well as on hopping between strands.

\subsection*{Freely diffusing tracer particles}

\begin{figure}[htbp] 
  \centering
  \includegraphics[width=0.95\linewidth]{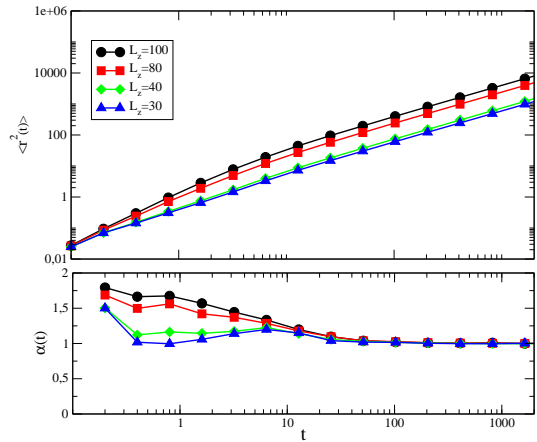}
  \caption{MSD and time-dependent exponent of the freely diffusion model
    tracer for different degrees of stretching of the polymer at
    interaction strength $\varepsilon=2$ and density $\rho=0.0084$.}
  \label{fig:m3_rho0_0084}
\end{figure}

\begin{figure}[htbp]
  \centering
  \includegraphics[width=0.95\linewidth]{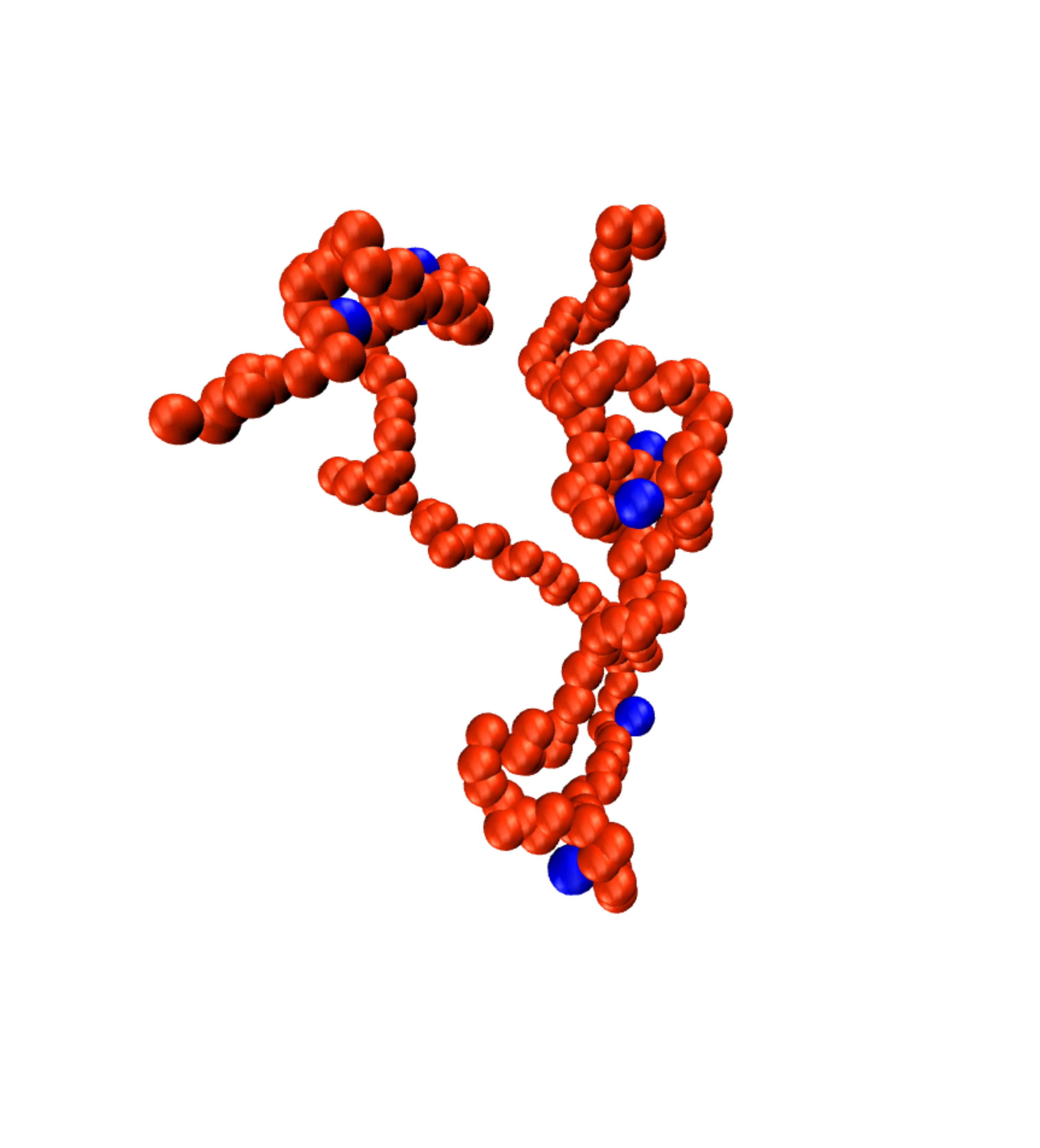}
  \caption{Simulation snapshot of the tracer particles getting stuck
    in the kinks of a coiled polymer.}
  \label{fig:m3_snap_kinks}
\end{figure}

Figure \ref{fig:m3_rho0_0084} depicts the MSD and time-dependent
exponent of freely diffusing tracer particles at different degrees of
stretching $L_z$ at a density of $\rho=0.0084$.  Comparing these plots
to the corresponding plots in the model with nearest-neighbor
interaction reveals a number of profound differences. The ballistic
behavior seen at short times disappears after even shorter times than
in the nearest-neighbor interaction model.  We think that the effect
is caused by the introduction of kinks into the model.  When employing
the usual all-neighbor Lennard-Jones interaction, the tracer feels the
deep potential wells along the polymer chain where the chain has a
kink, and can easily be trapped in them for long times (see Figure
\ref{fig:m3_snap_kinks}).  Since the potential well caused by the kink
is very narrow, it affects the ballistic behavior of the tracer
particle on very short time scales, even before the velocity
autocorrelation is destroyed by the Langevin thermostat.

Curiously, the plots of the time-dependent exponent at low density
still allows to distinguish between the different degrees of
stretching.  The reason for this is, that the number of kinks is
significantly larger for a coiled polymer chain than for a stretched
polymer chain.  Therefore, for very coiled polymer chains, $\alpha(t)$
drops much faster than for a stretched chain.  This is, however, not
the effect of the tracer particles sliding along the polymer contour
and shall not be further considered in this work.

The subdiffusive regime that was observed in the nearest-neighbor
model at similar densities has completely disappeared.  From the
ballistic regime at very short times, the time-dependent exponent
directly goes over into a normal diffusive regime.  We believe that in
these systems the anomalous diffusion caused by the tracer particles
sliding along the chain is on the one hand completely dominated by the
process of particles being immobilized by the kinks, and on the other
hand by the process of hopping, and therefore it is not visible in
these graphs.

\begin{figure}[htbp] 
  \includegraphics[width=0.95\linewidth]{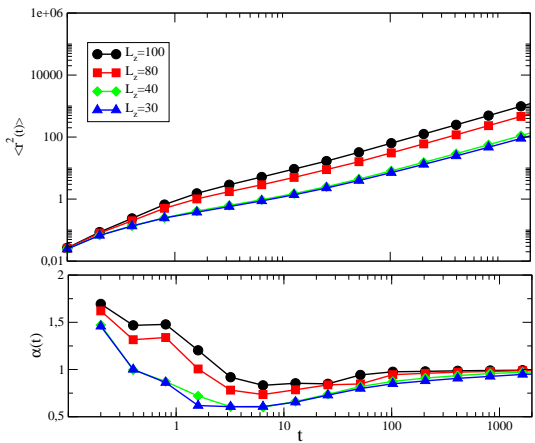}
  \caption{MSD and time-dependent exponent of the freely diffusion model
    tracer for different degrees of stretching of the polymer at
    interaction strength $\varepsilon=2$ and density $\rho=0.1$.}
  \label{fig:m3_rho0_1}
\end{figure}

To increase the importance of tracer particles sliding along the chain
over the process of hopping and sticking, we have increased the
density to significantly higher values of $\rho=0.1$, which should
decrease the importance of hopping.  The MSD and time-dependent
exponent are plotted in Figure \ref{fig:m3_rho0_1}.  In these plots,
the time-dependent exponent $\alpha(t)$ drops below $1$ in an
intermediate region, which might indicate subdiffusive behavior.
However, the subdiffusive behavior already sets in at very short
times, where the tracer particles have not yet diffused even a single
bead diameter.  Therefore at least this part of the subdiffusive
region can not originate in the tracer particles probing the chain
contour.  Instead, it can be assumed that it is a result of a
``hopping-between-cages''-process similar to the effect observed in
the random bead system in the nearest-neighbor interaction model, only
that in this case the cages are represented by the kinks rather than
by single beads.

Unfortunately, the plots do not allow to celarly decide whether
subdiffusive behavior caused by the tracers sliding along the polymer
contour is visible within the system. Since the densities used for
producing the data of these figures are very high, an experimental
verification of these features seem to be very improbable.

\section*{Conclusions}

In this work, we have investigated the effect of polymer strands (as
they might be found in a hydrogel) on the diffusion behavior of small
tracer particles (as they are used in FCS) that are attracted to the
polymer strands.  From theoretical considerations, we concluded that
when such tracer particles slide along the contour of a polymer
strand, they will exhibit subdiffusive behavior due to the fractal
nature of the polymer.  

To investigate this situation, we have devised a series of simplified
models with increasing complexity.  A simple numerical simulation of a
model with tracer particles that were strongly bound to frozen
conformations of a polymer strand proved the soundness of the
theoretical assumptions.  In plots of the mean square displacement
(MSD) and the time-dependent exponent, it was observed, that when the
polymer is stretched (\ie when the hydrogel is swollen), this affects
the structure of the polymer contour and the subdiffusive behavior of
the sliding tracer particles.  This is best rationalized in terms of 
Pincus blobs.  When the polymer strands are mostly relaxed, the blobs
are relatively large, and the subdiffusive behavior of the tracer
particles should be observable over many decades of the time scale,
while when the polymer strands are stretched, the Pincus blobs become
small and the subdiffusive behavior can only be observed on very short times.

In reality, tracer particles cannot be expected to be so strongly
bound to the polymer.  Instead, thermal kicks induced by the polymer
motion most probably will aid their desorption and particles can be
expected to diffuse freely between the polymer
strands, which is referred to as ``hopping''.  The influence of this
process was investigated by the means of a simple molecular dynamics
simulation model, where the tracer particles interacted with the
frozen polymer conformations via a nearest-neighbor Lennard-Jones
interaction.  We observed, that the hopping process easily dominates
the overall diffusion behavior, as it has a much higher diffusion rate
than the process of sliding along the polymer.  To be able to see the
subdiffusion caused by the tracer particles sliding along the polymer,
the most important parameter was determined to be the strength of the
attractive interaction between the tracer and the polymer chain, while
the density of the polymer strands was of minor importance.

In the last model, where the interaction between tracer particles and
the frozen polymer was modeled by a standard all-neighbor
Lennard-Jones interaction, we noticed that kinks within the polymer
chain have a profound impact on the diffusion behavior.  The
reason is the increased attraction of the tracer particles to the
kinks, which immobilizes the tracer particles and dominates the
diffusion behavior.  At low densities, the subdiffusive behavior
caused by the tracer particles sliding along the chain is completely
hidden.  Although at high density, an apparently intermediate
subdiffusive region appears, we noticed that it occurs at time and
length scales that are still within the direct vicinity of the
kinks.  Therefore we think it is caused by hopping between the kinks.

We believe that the large effect of these kinks on the diffusive
behavior is not as dominating in real hydrogel systems, but instead
can be seen as an artifact of our model which originates in the fact
that we use fixed polymer conformations in our model.  In the real
world, where the polymer is moving, potential wells caused by kinks
should be too short-lived to really catch tracer particles.

We should remark that the parameters of the simulational model systems
are applicable only to weakly crosslinked hydrogels, since normal
hydrogels have mesh sizes of about tens of nm.  This means that the
experimentally accessible length and time scales are larger than the
average network mesh size and hence the length of the pieces of
polymer strands that the tracer can slide along.  Experimentally, one
should be able to see the simulated anomalous effects only in weakly
cross-linked hydrogels that have large enough network mesh cells.

As final conclusion we believe that further studies are required to
determine whether it is possible to explore the polymer structure in a
hydrogel by the means of tracer diffusion studies.

\begin{acknowledgements}
  We acknowledge the financial support provided by the Deutsche
  Forschungsgemeinschaft (DFG) as part of the SPP 1259 ``Intelligente
  Hydrogele''.

  The authors want to thank Georg Fytas, Alexander Grosberg, Owen
  Hickey, Felix H\"{o}fling, Peter Ko\u{s}ovan, Ralf Metzler, Ricardo
  Raccis, and Roland Winkler for helpful discussions.
\end{acknowledgements}

\bibliographystyle{spphys}
\bibliography{paper}

\begin{thebibliography}{10}
\providecommand{\url}[1]{{#1}}
\providecommand{\urlprefix}{URL }
\expandafter\ifx\csname urlstyle\endcsname\relax
  \providecommand{\doi}[1]{DOI \discretionary{}{}{}#1}\else
  \providecommand{\doi}{DOI \discretionary{}{}{}\begingroup
  \urlstyle{rm}\Url}\fi

\bibitem{wink07a}
R.G. Winkler, J. Chem. Phys. \textbf{127}, 054904 (2007)

\bibitem{lumma03a}
D.~Lumma, S.~Keller, T.Vilgis, J.O. R{\"a}dler, Phys. Rev. Lett. \textbf{90},
  218301 (2003)

\bibitem{nandi08a}
C.K. Nandi, P.P. Parui, B.~Brutschy, T.L. Schmidt, A.~Heckel, Anal. Bioanal.
  Chem \textbf{390}, 1595 (2008)

\bibitem{bacia04a}
K.~Bacia, D.~Scherfeld, N.~Kahya, P.~Schwille, Biophys. J. \textbf{87}, 1034
  (2004)

\bibitem{cherdhirankorn09a}
T.~Cherdhirankorn, A.~Best, K.~Koynov, , K.~Peneva, K.~M{\"u}llen, G.~Fytas, J.
  Phys. Chem. B \textbf{113}, 3355 (2009)

\bibitem{cherdhirankorn09b}
T.~Cherdhirankorn, V.~Harmandaris, A.~Juhari, P.~Voudouris, G.~Fytas,
  K.~Kremer, K.~Koynov, Macromolecules \textbf{13}, 4858 (2009)

\bibitem{gallo03a}
P.~Gallo, M.~Rovere, J. Phys.: Condens. Matter \textbf{15}, 7625 (2003)

\bibitem{diet08a}
P.~Dieterich, R.~Klages, R.~Preuss, A.~Schwab, Proc. Natl. Acad. Sci.
  \textbf{105}, 259 (2008)

\bibitem{wong04a}
I.~Wong, M.~Gardel, D.~Reichman, E.~Weeks, M.~Valentine, A.~Bausch, D.~Weitz,
  Phys. Rev. Lett. \textbf{92}, 178101 (2004)

\bibitem{tolic04a}
I.M. Toli\'{c}-N{\o}rrelykke, E.L. Munteanu, G.~Thon, L.~Oddershede,
  K.~Berg-S{\o}rensen, Phys. Rev. Lett. \textbf{93}, 078102 (2004)

\bibitem{banks05a}
D.S. Banks, C.~Fradin, Biophys. J. \textbf{89}, 2960 (2005)

\bibitem{szymanski09a}
J.~Szymanski, M.~Weiss, Phys. Rev. Lett. \textbf{103}(3), 038102 (2009).
\newblock \doi{10.1103/PhysRevLett.103.038102}

\bibitem{saxton96a}
M.J. Saxton, Biophysical Journal \textbf{70}, 1250 (1996)

\bibitem{fytas08a}
G.~Fytas, Personal communication

\bibitem{sprakel07a}
J.~Sprakel, J.~van~der Gucht, M.A. {Cohen Stuart}, N.A.M. Besseling, Phys. Rev.
  Lett. \textbf{99}(20), 208301 (2007).
\newblock \doi{10.1103/PhysRevLett.99.208301}

\bibitem{hu06a}
T.~Hu, A.Y. Grosberg, B.I. Shklovskii, Biophys. J. \textbf{90}(8), 2731 (2006).
\newblock \doi{10.1529/biophysj.105.078162}

\bibitem{berg81a}
O.G. Berg, R.B. Winter, P.H. von Hippel, Biochemistry \textbf{20}, 6929 (1981)

\bibitem{flory53a}
P.J. Flory, \emph{Principles of Polymer Chemistry} (Cornell University Press,
  Ithaca, NY, 1953)

\bibitem{pincus76a}
P.~Pincus, Macromolecules \textbf{9}, 386 (1976).
\newblock \doi{10.1021/ma60051a002}

\bibitem{degennes79a}
P.G. de~Gennes, \emph{Scaling Concepts in Polymer Physics} (Cornell University
  Press, Ithaca, NY, 1979)

\bibitem{allen87a}
M.P. Allen, D.J. Tildesley, \emph{Computer Simulation of Liquids}, 1st edn.
\newblock Oxford Science Publications (Clarendon Press, Oxford, 1987)

\bibitem{padding06a}
J.T. Padding, A.A. Louis, Phys. Rev. E: Stat., Nonlinear, Soft Matter Phys.
  \textbf{74}(3), 031402 (2006).
\newblock \doi{10.1103/PhysRevE.74.031402}

\bibitem{milchev02a}
A.I. Milchev, in \emph{Computer simulations of surfaces and interfaces}, vol.
  114, ed. by A.I.M. Burkhard~D\"{u}nweg, David P.~Landau (Kluwer Academic
  Publishers, 2002), vol. 114

\bibitem{limbach06a}
H.J. Limbach, A.~Arnold, B.A. Mann, C.~Holm, Comp. Phys. Comm. \textbf{174}(9),
  704 (2006).
\newblock \doi{10.1016/j.cpc.2005.10.005}

\bibitem{benavraham00a}
D.~Ben-Avraham, S.~Havlin, \emph{Diffusion and reactions in fractals and
  disordered systems} (Cambridge University Press, 2000)

\bibitem{klages10a}
R.~Klages, in \emph{Reviews of Nonlinear Dynamics and Complexity}, vol.~3, ed.
  by H.G. Schuster (Wiley-VCH, 2010), vol.~3, pp. 169--216

\end{thebibliography}

\end{document}